\documentclass[a4paper,12pt]{elsarticle_rfabbri}

\usepackage[small,bf]{caption}
\usepackage{graphicx}
\usepackage[usenames,dvipsnames]{color}
\usepackage{amsfonts}
\usepackage{amssymb}
\usepackage{amsmath}
\usepackage{indentfirst}
\usepackage{natbib}
\usepackage{fancyhdr}
\usepackage[utf8]{inputenc}                              	  % CONVERSÃO EM PORTUGUÊS
\usepackage{bbm} 								%para escrever double 1
\usepackage{algorithm}						% para escrever algoritmos
\usepackage{algpseudocode}				%para escrever pseudo-códigos
\usepackage{subfigure} 							%poe varias figuras como se fossem uma só
\usepackage{url}	
\usepackage[bookmarks,pdfborder={0 0 0}]{hyperref} % cria índice com referencias no arquivo pdf, 
\usepackage[top=2.0cm, bottom=3.3cm, left=2.5cm,right=2.5cm]{geometry} % Margens
\usepackage{setspace}    					     % para colocar espaçamento entre linhas
\definecolor{seagreen}{RGB}{46,139,87}
\usepackage{scalefnt}
\usepackage{listings}

\newcommand{\keywordsname}{Keywords}

\newcommand{\code}[2]{
 \vspace{1em}
 \subsubsection*{#1}
 \lstinputlisting{#2}
}

\begin{document}

%\title{Modeling the Evolution of Fluid Surface Patterns from Video}
%\title{Video-based Modeling of Fluid Surface Patterns} 
%\title{Image-based Modeling of Fluid Surface Patterns} 
%\title{Modeling Fluid Surface Patterns from Video} 

%\title{A Video-based Fluid Surface Pattern Model} 
\begin{frontmatter}

\title{Visualization of EIS at large potential range -- new insights}

\author[iprj]{IVAN N.\ BASTOS\corref{corr1}}
\ead{inbastos@iprj.uerj.br}
\ead[url]{www.labcor.iprj.uerj.br}

\author[iprj]{MARCOS PAULO MOURA CARVALHO}

\author[iprj]{RICARDO FABBRI}

\author[cnrs]{RICARDO PEREIRA NOGUEIRA}

\address[iprj]{Instituto Polit\'{e}cnico, Universidade do Estado do Rio de
Janeiro\\C.P.: 97282 - 28601-970 - Nova Friburgo, RJ, Brazil}

\address[cnrs]{LEPMI UMR 5279 CNRS -- Grenoble INP -- Université de Savoie -- Université Joseph Fourier, BP 75, 38402 Saint Martin d’Hères, Institut National Polytechnique de Grenoble, INPG, France}

\cortext[corr1]{Corresponding Author.\  Fax: +55 22 2533 5149}

\begin{abstract}
Electrochemical Impedance Spectroscopy (EIS) is an experimental technique
largely used in electrochemistry and corrosion studies. However, almost all
published papers have just measured the EIS at the corrosion potential,
especially for corrosion purposes. This fact limits the capability of the
technique. In this paper, a Scilab software was developed which allows the
visualization of multiple EIS diagrams regarding the potential, exposure time or
experiment run. This procedure was applied to austenitic stainless steels in two
electrolytes from cathodic to anodic potentials. The EIS maps with two- or
three-dimensions were very useful to depict the evolution of the surface with
respect to the large range of applied potential. Some results are shown to highlight the
usefulness of this approach as a complementary technique to the DC test performed at
a given potential range.
\end{abstract}


\end{frontmatter}

\section{Introduction}

Electrochemical Impedance Spectroscopy (EIS) is a linear technique fundamental
to study the mechanisms of the electrochemical corrosion process. In this case, the
regulation of a direct potential or current in regions of the E-I plot to obtain
diagrams that reflects the surface status is
crucial~\cite{orazem:tribollet:book:2011}. Indeed, the intensity
of electrochemical phenomena is strongly dependent on the potential. Thus, the
scan of potential is an ordinary means to obtain the kinetic aspects of
interfaces. In this sense, polarization curves show the global stationary
response of an electrode, being, consequently, used in almost all corrosion
studies. However, these studies frequently measure the AC response (impedance)
only at the corrosion potential. Then, they have not correlated the aspects of
polarization curves with impedance diagrams for a given potential

The mapping procedure presented here is an attempt to fill
this gap for a broad potential range.  The mapping shown in this paper is
different from the dynamic EIS proposed by Darowicki and
co-workers~\cite{darowicki:etal:ElectrochimicaActa2004}. Their
measurement uses a package of few superimposed sine waves with direct potential.
The direct potential sweeps at a relatively high rate, therefore the impedance
spectra are determined for narrow periods of time. Thus, even a non-stationary
process could be measured. On the other hand, the  measurement procedure used in
the present work is the classic one~\cite{oltra:keddam:corrosionScience1988}, measuring on a frequency by frequency
basis. However, we stressed the advantages for data interpretation when the full spectrum
is continuously available for all potential, as present in the polarization curves.
Therefore, we believe that this improved technique can be of interest to users,
as it was the case of the reference~\cite{vieira:etal:JMRT2013} related to optical microscopy, because
it allows a better understanding of EIS behavior performed at potential range.

\section{Experimental Procedure}

Samples of UNS S30400 (304SS) and S31600 (316SS) stainless steels were used in
the as-received microstructures. The chemical composition of the steels is shown
in Table~\ref{tab:chem:composition}.

\begin{table}
\centering
\caption{Chemical composition of stainless steels $(\% wt )$}
\label{tab:chem:composition}
	\begin{tabular}{|c||c c c c c c c c c c|}\hline
\textbf{Steel}		& \textbf{C} & \textbf{Cr} 
& \textbf{Cu} & \textbf{Mn} & \textbf{Mo} & \textbf{N} & \textbf{Ni} &
\textbf{P} & \textbf{S} & \textbf{Si}\\ \hline
304SS &0.038&18.21 &0.49 &1.47 & 0.28 &0.079 & 8.12 & 0.037 & 0.025 & 0.5\\
316SS &0.020&16.55 &0.38& 1.81 &2.09 &0.073 & 10.03 & 0.034 & 0.024 & 0.3 \\\hline
%\\ \hline\hline 
%    \rule[-1ex]{0pt}{2.5ex} \textbf{Group 1} & 0.1099 & 0.003 \\  
%    \rule[-1ex]{0pt}{2.5ex} \textbf{Group 2} & 0.1105 & 0.039 \\  
%    \rule[-1ex]{0pt}{2.5ex} \textbf{Group 3} & 0.1064 & 0.056 \\  
%    \rule[-1ex]{0pt}{2.5ex} \textbf{Group 4} & 0.1092 & 0.005  \\\hline
%		\rule[-1ex]{0pt}{2.5ex} Group 1 & 0.1099184 & 0.0032635 \\ \hline 
%		\rule[-1ex]{0pt}{2.5ex} Group 2 & 0.1105400 & 0.0390918 \\ \hline 
%		\rule[-1ex]{0pt}{2.5ex} Group 3 & 0.1064335 & 0.0562951 \\ \hline 
%		\rule[-1ex]{0pt}{2.5ex} Group 4 & 0.1092023 & 0.0056376  \\ \hline 
	\end{tabular} 
\end{table}

Two solutions were employed: one $3.5\%\, NaCl$; and another $NaHCO_3$
($0.025\, mol\cdot L^{-1}$),
$Na_2 CO_3$ $(0.025\, mol\cdot L^{-1})$ and $NaCl$ ($0.60\, mol\cdot L^{-1}$).
The latter solution has a $pH$
close to 10. They mimic the seawater and the pore solution of a carbonated
concrete with chloride infiltration, respectively. All tests were carried out in
aerated electrolyte and under controlled temperature $(25.0\pm 0.2)$. 

\begin{figure}[h!]
\centering
\includegraphics[width=0.5\textwidth]{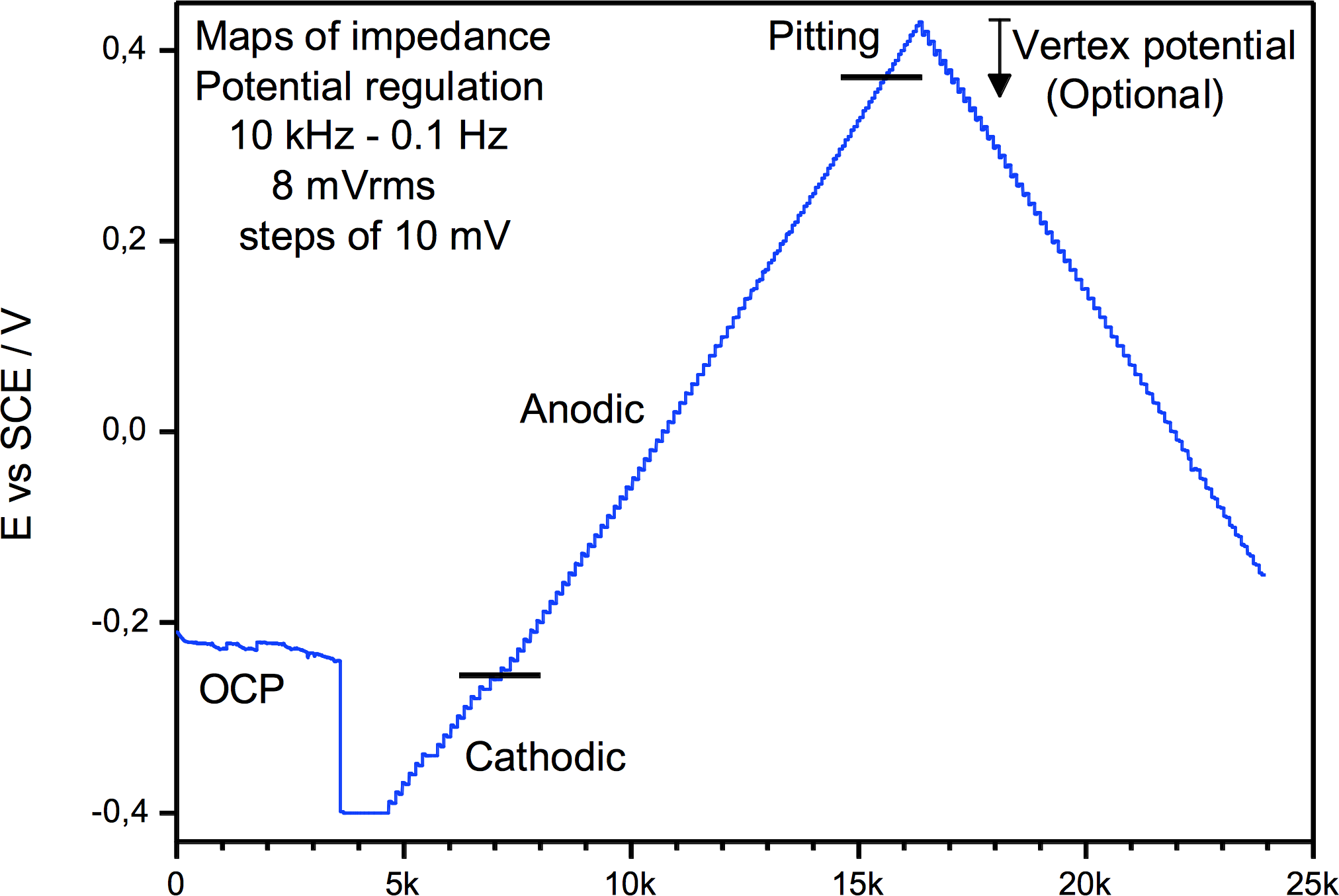}%
\caption{% 
DC potential scan used to evaluate the impedances.
}\label{fig:ivan:1}
\end{figure}

The counter-electrode was a platinum grid and a saturated calomel electrode
(SCE) was employed as a reference electrode. An open circuit potential (OCP) of one hour
was held to allow the steady-state condition to be reached. The staircase
potentiostatic polarization curves were carried out starting at $-400 mV$ up to
$400\, mV \times SCE$ or up to a specific potential. Steps of $10 mV$ were used to
perform this scanning. After $100 s$ stabilization time at the new potential, the
impedance was measured. Fig.~\ref{fig:ivan:1} shows the complete potential cycle, i.e, one hour
of open circuit, a potential scan from cathodic to anodic region. Eventually, a
reversion of potential can be done at vertex point. The average potential rate
during the staircase-like scan is $70 \mu V\cdot s^{-1}$, low enough to permit a steady-state
measure of impedance performed at a relatively high frequency $(f \geq 0.10 Hz)$. If
a lower frequency is used, the potential scan rate decreases. In any case,
the mapping is obtained at a rate below that of the ordinary polarization curve.

Due to the vast amount of data from the hundreds of diagrams (a typical
worksheet has around ten thousand points) specialized software was developed to
plot the maps. A software application called EIS-Map was written in the Scilab 5.4.1
language~\cite{scilab,Fabbri:etal:Arxiv2012} to enable the complete visualization and post-processing of the 2D and 3D impedance
spectra in Bode format. Commercial graphing software do not handle such a large amount of data. EIS-Map
was used to generate the maps of phase angle and magnitude of the impedance of
the desired samples. The EIS-Map software can be downloaded from [6]. A Section of
the code to plot the impedance maps is shown below.

\code{Section of the Scilab code to plot impedance maps}{eis-plot-snippet.sci}

It was then possible to render the 2D and 3D plots in a smooth way due to the
parameters of the experiment – steps of $10 mV$ and perturbation of $8 mV$ -- which
enable an even superposition of the diagrams. Moreover, the \texttt{plot3d1} Scilab
function performs an interpolation of consecutive diagrams.\\[0.5cm]

\section{Results and discussion}

\begin{figure}
\centering
\footnotesize{\textbf{a}}\includegraphics[width=0.49\textwidth]{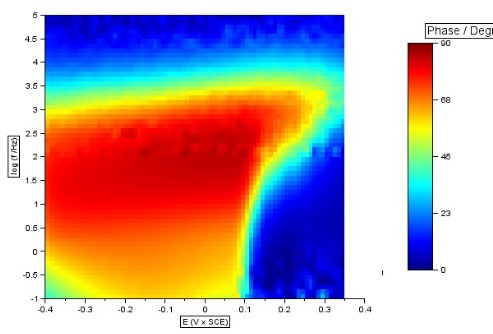}%
\ \ \ \footnotesize{\textbf{b}}\includegraphics[width=0.455\textwidth]{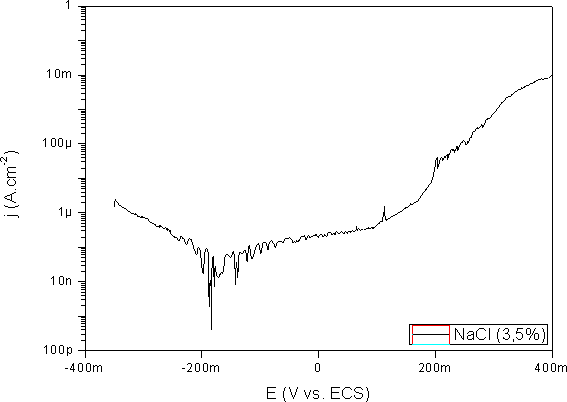}%
\caption{% 
a) Phase angle of 304SS, b) the related I-E plot. Pitting onset occurred
close to $0.10\, V \times SCE$
}\label{fig:ivan:2}
\end{figure}

A comprehensive dependency of impedance diagrams on potential can be determined
by mapping the magnitude and phase angle. Thus, 75 diagrams were performed
consecutively, each one being separated by $10 mV$ from the previous ones, as shown in
Fig.~\ref{fig:ivan:2}a. The actual E-I plot is depicted in Fig. 2b. This plot is not the
polarization curve because it was slightly affected by the sine wave
perturbation of $8 mV$, in both potential and the related current data. It nevertheless informs the
DC aspects of electrode, such as the cathodic domain, anodic and the pitting. 

\begin{figure}
\centering
\footnotesize{\textbf{a}}\includegraphics[width=0.51\textwidth]{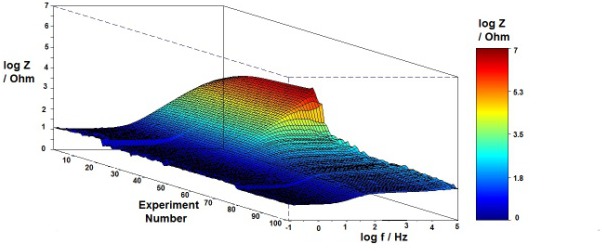}%
\footnotesize{\textbf{b}}\includegraphics[width=0.48\textwidth]{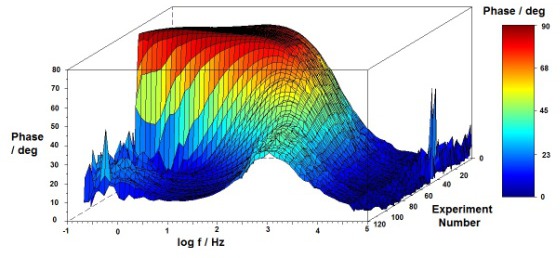}%
\caption{% 
a) 3D map of EIS modulus, b) Phase angle. Solution $3.5\%\, NaCl$. The applied DC
potential increases with experiment number.
}\label{fig:ivan:3}
\end{figure}

The corrosion potential of Figs.~\ref{fig:ivan:2} and~\ref{fig:ivan:3} is approximately $-0.2\, V \times SCE$.
The frequency spectrum of high and medium phase angle enlarges with the
applied potential. The films formed at anodic potential show better
properties as the potential increase, from cathodic to anodic region.
Accordingly, the EIS modules behaves similarly. This aspect can be seen on
higher modulus as well as a higher angle for a wide frequency span. High
angles and moduli are depicted in red color in the maps. Thus, the anodic
potential increases and the film quality also increases, even in relation to
corrosion potential. However, this applied potential increases the
driven-force for the film breakdown. On the other hand, low potential is related to
cathodic process intensity since the steels do not corrode, and the impedance
depicts the intensity of the cathodic process that takes place on the electrode
surface. Beyond the corrosion potential, the surface is under anodic regime
and an observable improvement of the film is noted. The frequency range where
the angle closer to $90^\circ$ (red) also alters with potential, from 
$10^{0.5}-10^3 (3-1,000) Hz$ at the corrosion potential to $10^{0.5}-10^{3.5} (3-2,500) Hz$ just
before the pitting. The spatial localization of corrosion sites also localizes the
characteristic frequency range.

Interestingly, the sudden contraction of high-medium (red-yellow) at 
$100\, mV \times SCE$ is more sensible to detect the onset of pitting than the current of
Fig.~\ref{fig:ivan:2}b that changes  smoothly. The sudden decay of the magnitude, whereas the
angle at low frequency approaches a resistive behavior, means that the angle
at low frequency approaches zero. The characteristic frequency changes to
circa $1 kHz$ and the maximum angle is around $50^\circ$. Moreover, the previous
widespread angle distribution, with a true plateau at low frequency, reduces
sharply in high potential (Fig.~\ref{fig:ivan:3}b located around $1\, kHz$). In
Fig.~\ref{fig:ivan:3} the
diagrams are displayed in sequence of diagrams instead of potential.

\begin{figure}[htb!]
\centering
\includegraphics[width=0.5\textwidth]{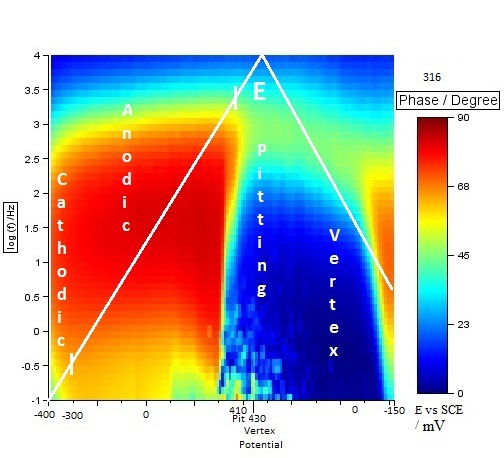}%
\caption{% 
Electrochemical impedance maps of 316SS with reversion of potential. Beyond
$+0.430\, V \times ECS$ there is a reversion of potential scan up to
$-0.15\, V \times SCE$. $pH$ close to 10.
}\label{fig:ivan:4}
\end{figure}

The reversion of potential occurred at $-0.43\, V \times SCE$, after a pitting
breakdown. This type of mapping allows a complete visualization of impedance for
a large potential range. After the pitting, the modulus decays well below $1
kOhm$ and it takes at least $200 mV$ to  recover high values (c.a. $60^\circ$).
The pits are a physical cavity as can be seen in Fig.~\ref{fig:ivan:4}.
Moreover, they are
electrochemically depressed at a lower potential. A subtle arc (yellow color) from position $(E=410\,
mV \times SCE,\ \log f = 3)$ to $(E=50\, mV \times SCE, \log f = 2)$ is an
additional track of repassivation, not available in DC measurement or a single
impedance diagram. 

Albeit a strong angle scattering near the pitting potential, capacitive angles
with low angle at a characteristic frequency around $400 Hz$ occurs. With the
potential reduction, the properties improve again, however, under a
discontinuous manner and always with values below those of the corresponding
potential of forward scan. For instance, the impedance at $-150\, mV \times SCE$ after
vertex is lower (light yellow, low frequency) than $-150\, mV \times SCE$ at anodic
region (dark red, forward).  Fig.~\ref{fig:ivan:4} shows the optical observation of pitting
attack of 304SS after a complete EIS mapping in 3.5\% NaCl. 

\begin{figure}
\centering
\includegraphics[width=0.5\textwidth]{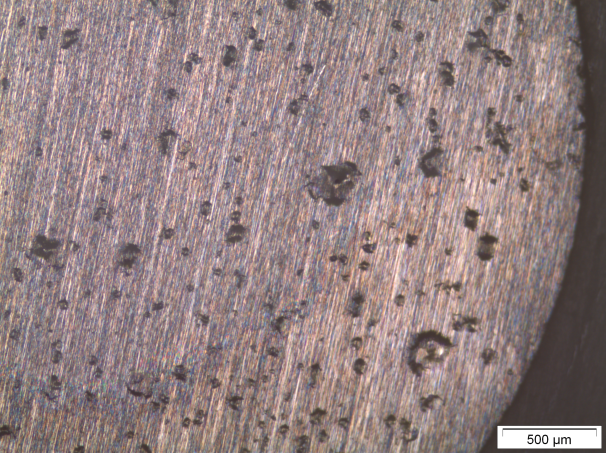}%
\caption{% 
Pitting observed after the EIS mapping of 304SS in $3.5\% NaCl$ in optical microscopy.
}\label{fig:ivan:5}
\end{figure}

\section{Summary}

Multiple electrochemical impedances were performed under continuous mode. The
potential scan rate was well bellow the usual value in polarization curves,
thus a quasi-stationary condition was assumed. Several diagrams were obtained
and a special software was implemented to allow a proper visualization of these
diagrams,
in two and three dimensions. As examples of advantages, the continuous
improvement of film in stainless steels corresponding to the potential up to the pitting
potential is clearly depicted. Moreover, the inhibition of anodic activity, in
the present case related to pitting, can be tracked by the visual impedance
display, not available in DC evaluation nor in a single impedance diagram.
Thus, by using EIS maps, the flat phase angle exhibited by stainless steels
close to corrosion potential changes to a  compressed peak at high frequency when
pitting is present.

\section*{Acknowledgements}
The authors thank the Brazilian agencies: CNPq, CAPES, and FAPERJ for the financial support.

\bibliographystyle{elsarticle-num}
\bibliography{materials,personal}

\end{document}